  \documentclass[preprint]{aastex}
  \usepackage{natbib}
  \usepackage{emulateapj5}
  \usepackage{graphicx}
\newcommand{\msb}{ $\left<\mu\right>$}
\newcommand{\cgm}{$C$ vs. $G$ vs. $\left<\mu\right>$}
\newcommand{\gb}{$g^*$-band}
\newcommand{\ib}{$i^*$-band}
\newcommand{\samplesize}{930~}

\begin{document}

\bibliographystyle{apj}


\title{A New Approach to Galaxy Morphology: I. Analysis of the {\em Sloan Digital Sky Survey Early Data Release}}
\author{Roberto G. Abraham}
\affil{
   Department of Astronomy \& Astrophysics,
   University of Toronto, 60 St. George Street,
   Toronto, ON, M5S~3H8, Canada.\\
   abraham@astro.utoronto.ca}
\medskip
\author{Sidney van den Bergh}
\affil{
   Dominion Astrophysical Observatory,
   Herzberg Institute of Astrophysics,
   National Research Council,
   5071 West Saanich Road,
   Victoria, British Columbia,
   V9E~2E7, Canada.\\
   sidney.vandenbergh@nrc-cnrc.gc.ca}
   
\medskip

\author{Preethi Nair}
\affil{
   Department of Astronomy \& Astrophysics,
   University of Toronto, 60 St. George Street,
   Toronto, ON, M5S~3H8, Canada.\\
   nair@astro.utoronto.ca\\}

\begin{abstract}
In this paper we present a new statistic for quantifying galaxy morphology based on measurements of the Gini coefficient of galaxy light distributions. This statistic is easy to measure and is commonly used in econometrics to measure how wealth is distributed in human populations. When applied to galaxy images, the Gini coefficient provides a quantitative measure of the inequality with which a galaxy's light is distributed amongst its constituent pixels.  We measure the Gini coefficient of local galaxies in the {\em Early Data Release} of the {\em Sloan Digital Sky Survey} and demonstrate that this quantity is closely correlated with measurements of central concentration, but with significant scatter. This scatter is almost entirely due to variations in the mean surface brightness of galaxies. By exploring the distribution of galaxies in the three-dimensional parameter space defined by the Gini coefficient, central concentration, and mean surface brightness, we show that all nearby galaxies lie on a well-defined two-dimensional surface (a slightly warped plane) embedded within a three-dimensional parameter space. By associating each galaxy sample with the equation of this plane, we can encode the morphological composition of the entire SDSS \gb~sample using the following three numbers: \{22.451, 5.366, 7.010\}. The \ib~sample is encoded as: \{22.149, 5.373, and 7.627\}.
\end{abstract}


\section{INTRODUCTION}
\label{sec:introduction}

In 1926 Hubble introduced his ``tuning fork" classification system, which has proved to be admirably suited to classification of the vast majority of nearby bright field galaxies (Hubble 1926). Since all of the galaxy prototypes used by Hubble (1926,1936) are luminous giants or supergiants it is not surprising that his classification system is not well suited to the classification of low-luminosity galaxies (van den Bergh 1960ab; Sandage et al. 1987, 1994). Furthermore Hubble's classification system loses resolution in the cores of rich clusters, where almost all galaxies belong to classes E, S0 or SB0. Other types of problems arises when one attempts to extend the Hubble classification system to galaxies that are situated at large look-back times. In the first place the dichotomy between normal and barred spirals, which lies at the heart of the Hubble tuning fork classification system, appears to break down at large look-back times (van den Bergh et al. 1996, 2002; Abraham et al. 1999) as bars become increasingly rare at large $z$. Secondly, the fraction of peculiar galaxies, {\em i.e.} objects that do not fit comfortably within the Hubble scheme increases precipitously with increasing redshift (Griffiths et al. 1994;  Cowie, Hu \& Songaila 1995; Glazebrook et al. 1995; Driver, Windhorst \& Griffiths 1995;  Abraham et al. 1996, 1999; van den Bergh et al. 1996, 2000, 2001; Brinchmann et al. 1998; Driver et al. 1998; Marleau \& Simard 1998; Dickinson 1999; Dickinson et al. 2000. See Ellis 2001 and Abraham \& van den Bergh 2001, 2002 for recent reviews). In view of these difficulties it appears desirable to search for more general types of classification systems that might be applicable to dwarf and giant galaxies, to both field and cluster galaxies, and to objects at large redshifts. 

A first attempt to formulate such a very general classification system was made by Morgan (1958,1959) and codified in the Yerkes system. In the Yerkes system galaxies are classified primarily on the basis of their central concentration of light on the sequence $a-af-f-fg-g-gk-k$, where objects of type $k$ have the highest central concentration of light and late-type integrated spectra, whereas galaxies of type $a$ have a low central concentration of light and early-type integrated spectra. By design, Morgan's system tracks the principal correlation internal to the Hubble sequence, namely that between bulge-to-disk ratio and integrated color. 

The Yerkes system has never been widely adopted, but the fundamental idea behind it has recently undergone a renaissance. The basis for Morgan's system is the estimation of a single parameter (central concentration of light), that is both fully quantifiable (using techniques unavailable to Morgan, who estimated concentration by eye) and robust at low signal-to-noise levels. Both of these qualities are highly desirable in morphological investigations (particularly those of  distant galaxies based on data from the {\em Hubble Space Telescope}). A number of studies have therefore used central concentration as a {\em de facto} stand-in measure for bulge-to-disk ratio, and, by inference, a measure of position on the Hubble sequence (Doi et al. 1993; Abraham et al. 1994, 1996, 1999; Brinchmann et al. 1998;  Glazebrook et al. 1998; Takamiya 1999; Menanteau et al. 1999; 2000; Bershady, Jangren \& Conselice 2000; Volonteri et al. 2000; Trujillo et al. 2001; Kuchinski et al. 2001; Corbin et al. 2001; Shimasaku et al. 2001). Concentration measures are also used as a tracer of morphology in the  {\em Sloan Digital Sky Survey} (SDSS) database. (We refer the reader to York et al. 2000 for an overview of the SDSS). A comparison of SDSS concentration measures to  visual morphological classifications is given in Shimasaku et al. (2001), who find them well-correlated with Hubble type. And in what might (in some respects) be considered a modern re-appraisal of the fundamental basis of the Yerkes system, Kauffmann et al. (2002) has  shown that central concentration is strongly correlated with luminosity-weighted stellar age and star-formation history for 80,000 galaxies in the SDSS. 

In both the low-redshift and high-redshift contexts we have just described, measurements of central concentration have been based on simple aperture photometry. Such concentration index measurements therefore rely upon two key assumptions: Firstly, the measurements depend upon an assumed symmetry in the galaxies. (In the case of concentration measures based on elliptical apertures it is assumed that galaxy isophotes can be well described by ellipses, while in studies which use circular apertures it is assumed that inclination effects are small.) The second assumption that underlies aperture-based concentration index measurements is that galaxy images have a well-defined center. Inspection of the images of high-redshift galaxies (many of which are strongly asymmetric and distorted) shows that neither of these assumptions is likely to be fulfilled when studying galaxies in the distant Universe. This has lead us to seek out an alternative, more general, formulation of galaxy concentration that does not involve any kind of aperture photometry.   

In \S\ref{sec:gini} we present our proposed alternative definition of galaxy concentration, based on measurement of the Gini coefficient (Gini 1912), a familiar tools of econometrics which does not appear to have been used in an astronomical context before now.  We describe how the Gini coefficient can be viewed as a generalized measure of concentration that is applicable to galaxies of arbitrary shape, and which does not even require that a galaxy image have a single well-defined nucleus. We show that the Gini coefficient can be used to describe the morphology of all galaxy types, from perfectly symmetric nearby objects to the most distant, most wildly distorted images of strongly gravitational-lensed background galaxies.  

In \S\ref{sec:sdss} we measure the Gini coefficient for a magnitude-limited sample of \samplesize  $g^* < 16$ mag galaxies taken from the SDSS\footnote{We follow the convention of using asterisks following filter names to indicate that magnitudes are given in the preliminary SDSS AB system. The reader is referred to Fukugita et al. (1996), Hogg et al. (2001) and  Smith et al. (2002) for a discussion of the SDSS filter system and its calibration.}, and explore the relationship between the Gini coefficient and central concentration. We will show that these parameters are strongly correlated, but that they do not measure precisely the same thing. In cases where it is meaningful to measure {\em both} the Gini coefficient and central concentration (e.g. in nearby undistorted systems with well-defined centers), the difference between these two quantities is strongly correlated with mean surface brightness. Even more surprising is the fact that {\em all galaxies in the nearby Universe (irrespective of age, mass, star-formation history, dynamical state or morphology) fall on a two-dimensional surface within the three dimensional parameter space defined by the Gini coefficient, mean surface brightness, and central concentration}.  The implications of this are are considered in \S\ref{sec:discussion}, where we also describe our plans for future papers in this series. Our main conclusions are summarized in \S\ref{sec:conc}.


\section{THE GINI COEFFICIENT}
\label{sec:gini}  

The  Lorenz curve  (Lorenz 1905) is commonly used in economics to describe the inequality in a population's distribution of wealth. In this context the curve is constructed by plotting the cumulative proportion of income as a function of population rank. In a population where all individuals have exactly the same income (where, for example, $20$\% of the population has $20$\% of a country's total wealth), the Lorenz curve is a straight diagonal line with a slope of unity, called the line of equality. If there is any inequality in income, then the Lorenz curve falls below the line of equality. For example, in the extreme case where a tiny proportion of the population has nearly all the income, the Lorenz curve is flat and near zero for most of its length until rising precipitously near its end (see Figure 1).

\begin{figure*}[htb]
    \centering
    \includegraphics[width=4in,angle=0]{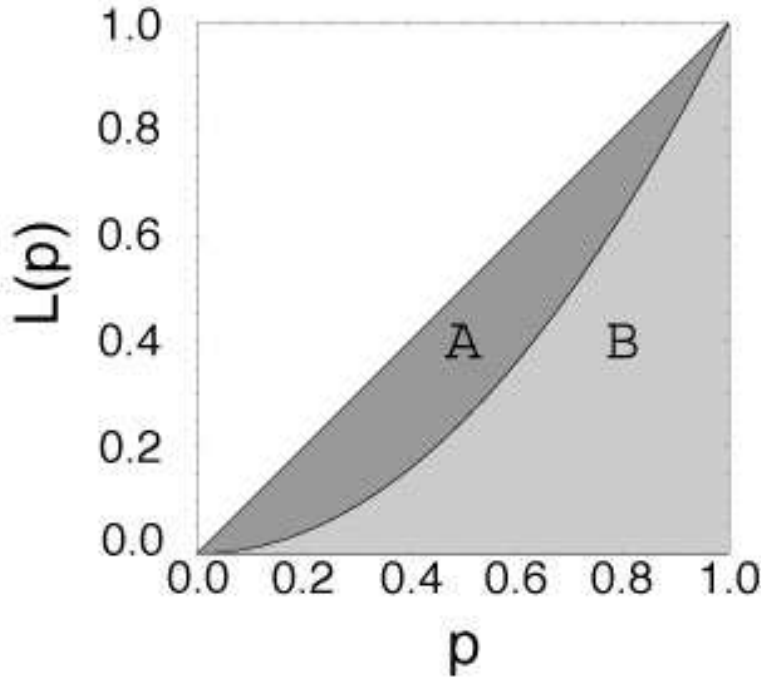}
    \caption{The geometric interpretation of the Gini coefficient based on the Lorenz Curve. The x-axis corresponds to the quantile of the distribution, the y-axis corresponds to the cumulative proportion. The Lorenz curve for a perfectly equal distribution corresponds to the diagonal line of equality.  In the figure, a schematic Lorenz curve divides the area beneath the line of equality into two area, A and B. The greater the deviation of a measured Lorenz curve from the line of equality, the greater the inequality. The Gini coefficient corresponds to the ratio of area A to the total area under the diagonal A+B.}
    \label{fig:GiniSchematic}
\end{figure*}  

The total amount of inequality  is conveniently parameterized using a summary statistic of the Lorenz curve, the Gini coefficient, $G$. The geometric meaning of $G$ is illustrated in Figure~1. The Gini coefficient is simply the ratio between area {\em A}  (the area enclosed between the line of equality and the Lorenz curve), to the total triangular area under the line of equality (A+B). The Gini coefficient ranges from a minimum value of zero, when all individuals are equal, to a maximum of one in a population where all the wealth is concentrated in a single individual\footnote{In other words, $G = 0$ for a perfect communist society and $G = 1$ for an absolute monarchy where all riches belong to the king!}. 

A more formal statistical description of the Lorenz curve and Gini coefficient complements the  intuitive description given above, by highlighting the close connection between the Gini coefficient and the absolute mean difference in a population, and by showing how the statistic can be computed with trivial computational cost. Let $X$ be a positive random variable with a cumulative distribution function $F(x)$, and let $X_i$ denote one of $n$ random deviates drawn from $X$. The Lorenz curve is then given by:

\begin{equation} 
L(p) = {1\over \bar{X}} \int_0^p F^{-1}(u)\ du. 
\end{equation}

\noindent The Gini coefficient is defined as the mean of the absolute difference between all combinations of $X_i$:

\begin{equation} 
G =  {1\over 2 \bar{X}n(n-1)} \sum_{i=1}^n \sum_{j=1}^n | X_i - X_j | 
\end{equation}  

\noindent where $\bar{X}$ is the mean value over all $X_i$. Numerical tests have shown that the estimated error on a measurement of $G$  can be reliably obtained using a statistical bootstrap (Efron \& Tibshirani 1993; Dixon et al. 1987), so it is desirable to use a faster algorithm for calculating $G$ than is provided by this formula. A very efficient way to calculate $G$ (Glasser 1962) is to first sort the $X_i$  into increasing order and then do a simple summation: 

\begin{equation} 
G = {1\over  \bar{X} n(n-1)} \sum_{i=1}^n (2 i - n -1) X_i\ \ \ \ \ \ \ \ (n>2).
\end{equation}  

\noindent The computational cost of calculating $G$ is therefore dominated by the computational cost of a single one-dimensional sort (which rises as $n \log n$ for an efficient sorting algorithm). 

\begin{figure*}[htb]
    \centering
    \includegraphics[width=6.5in,angle=0]{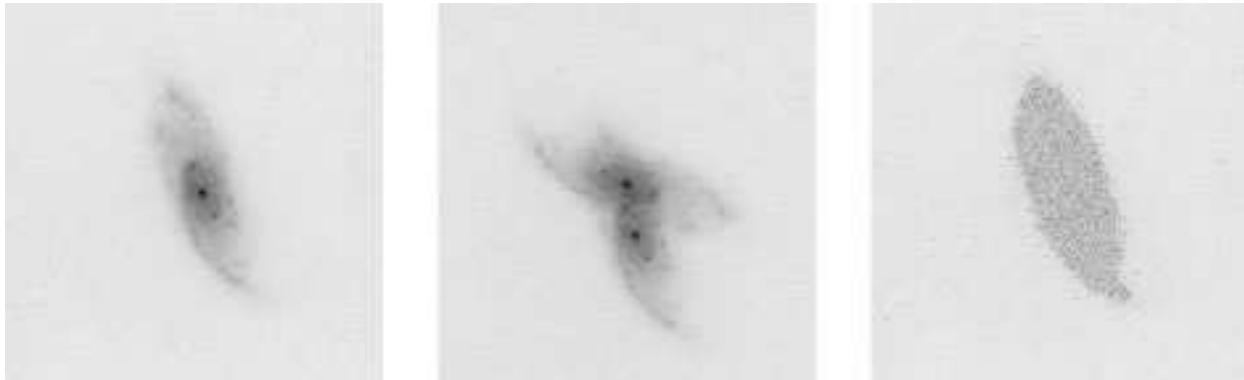}
    \caption{Three systems with essentially the same Gini coefficient but very different central concentrations. (Left) An $R$-band image of NGC 2715 from the digital catalog of Frei et al. (1996). (Middle) An image obtained by jackknife replication of the original NGC~2715 image. Since each pixel in the galaxy's image is simply a permutation of the original pixel set, $G$ remains unchanged. (Right) An image constructed by co-adding the original image of NGC 2715 with a rotated and displaced version of itself. Except within the region of overlap the shape of the pixel intensity distribution is preserved, so $G$ remains nearly unchanged. }
    \label{fig:GIsNotC}
\end{figure*}  

In the context of galaxy morphology, $G$ is calculated by applying Equation (3) to a list of pixel values sorted by intensity. Can a connection then be made between measures of $G$ as defined above and more conventional morphological parameters? To a first approximation, treating $G$ as a sort of generalized concentration index is not too far off the mark.  As shown in the next section, there  is a strong correlation between the Gini coefficient and central concentration for local galaxies. This suggests that $G$ might be a suitable replacement for concentration measures in studies of high-redshift galaxies, as a large fraction of these objects are distorted and peculiar. Measurements of $G$ are completely independent of galaxy shape, do not rely on any kind of aperture photometry, and do not even require that the galaxy image have a single well-defined center. (The implications of the fact that many distant galaxy images do not have a well defined centers will be investigated in Paper II of this series). However, the use of $G$ as a straightforward substitute for concentration indices is something of an oversimplification, at least for nearby galaxies where the fraction of peculiar systems is small. As illustrated in Figure~\ref{fig:GIsNotC}, it is easy to show that galaxy images can be transformed in many ways which preserve $G$ while completely changing measures of concentration. Clearly these two quantities do {\em not} simply measure  the same thing, and $G$ captures something fundamentally new in the image that is not probed by concentration measures alone.

Finally, we conclude this section by noting that the Gini coefficient may prove of general interest to astronomers in a number of areas unrelated to galaxy morphology. As a robust measure of dispersion that is insensitive to outliers and constrained to the unit interval, we suggest that it might prove a useful replacement for the variance in quantifying the spread of distributions that are strongly non-gaussian


\section{APPLICATION TO THE SLOAN DIGITAL SKY SURVEY}
\label{sec:sdss}

\subsection{Sample Definition}

Measurements of the Gini coefficient were made from the $g^*$-band and $i^*$-band ``Atlas'' images of all \samplesize galaxies with known redshifts and Petrosian magnitudes $g^*<16$~mag in the {\em Early Data Release} (EDR) of the {\em Sloan Digital Sky Survey} (SDSS).  We refer the reader to Stoughton et al. (2002)  for a description of the SDSS EDR, and simply note here that the Atlas images are small calibrated ``postage stamp'' images of individual galaxies in the EDR database.  The galaxies investigated span a broad range of morphological types and luminosities.  Measurements were made using software written by one of us (RGA) which determines $G$, as well as central concentration $C$, and mean surface brightness \msb, and many other photometric and morphological parameters.  $C$ is determined using the procedure outlined in Abraham et al. (1994; 1996). In this procedure second-order image moments are used to define a best-fitting outer elliptical aperture whose area is equivalent to that of the galaxy above an isophotal threshold. A smaller inner aperture is also defined with the same shape but with a semi-major axis length that is 30\% of that of the outer ellipse. The concentration index  is then given as the ratio of the flux in the inner and outer apertures. The mean surface brightness  \msb~ is determined in the most straightforward manner possible by simply dividing the total galaxy flux above an isophotal treshold over the galaxy area enclosed within the same threshold. The same isophotal threshold was used for measurements of $C$, $G$, and \msb, and the isophotal level chosen for each filter is described below.

We decided to forego analysis of the SDSS $u^*$-band and $z^*$-band data, which are shallow. For simplicity we will not discuss measurements obtained from the $r^*$-band Atlas images in the present paper, because the salient trends obtained with this filter are are simply intermediate between the $g^*$ and $i^*$-band trends presented here. A summary of measurements obtained for each object is presented in Table~1 (an electronic supplement to this paper). We have assumed $H_0=70$ km~sec$^{-1}$~Mpc$^{-1}$, $\Omega_M=0.3$, and $\Omega_\Lambda=0.7$ when computing the distance moduli and absolute magnitudes in this paper.

Each SDSS Atlas image encompasses only a single object and is ``segmented'' from the sky background to retain only those pixels assigned to the galaxy\footnote{In the lexicon of image processing ``segmentation'' refers to the process in which pixels are tagged as belonging to the galaxy or to the sky or to superposed objects. The EDR Atlas images are segmented by the SDSS Photo  image processing pipeline (Lupton et al. 2001), retaining data values only for pixels assigned to the galaxy. The remaining pixels in the Atlas image are assigned a uniform pedestal value of 1000 ADU.}. The images have been pre-processed to remove foreground stars and other contaminants. Our magnitude limit of  $g^*=16$ mag was chosen on the basis of visual inspection of the galaxy images, and represents what is in our view the limit beyond which detailed morphological work becomes challenging in  the SDSS data. Our impression is that at $g^*=16$ mag reliable Hubble types can only be assigned within very broad bins, and luminosity classifications on the basis of spiral structure is impossible, but simple classification into broad early-vs-late or spiral-vs-elliptical-vs-peculiar bins remains relatively straightforward. 

We now turn to our choice of isophotal threshold level used in our measurements of $C$, $G$, and \msb. Images from the SDSS EDR Atlas dataset are already segmented from the sky, so it is worthwhile to first consider why an additional thresholding step is needed. The SDSS photometric pipeline is optimized for the processing of galaxies somewhat fainter than those studied in this paper.  Some Atlas images of bright galaxies exhibit artifacts which prevent direct use of the images without additional pre-processing to better extract the galaxy from the background sky. The main problem is that in many cases too much sky is assigned to the galaxy images. This problem manifests itself as poor sky definition; for example long streaks of sky sometimes dominate the area of the some Atlas frames, leaving the galaxy lying in a corner of the postage stamp image. We estimate that problems of this sort occur in at least 15\% of the images. Small errors such as these result in little overall additional light being assigned to the galaxy (and thus they make negligible difference to inferred total magnitudes), but by inflating measured galaxy areas they can skew morphological measurements. In addition, separating galaxy pixels from sky pixels by using a cut at a pre-defined isophotal threshold is a fairly standard part of most morphological investigations, so the small amount of post-processing we have done might also be viewed as simply a standardization of the Atlas data which enables us to more easily compare them with previous work\footnote{It is worth noting that we were unable to locate photometric zeropoints for each Atlas image in the SDSS EDR database, and consequently these were ``reverse engineered'' using the Petrosian radii and Petrosian magnitudes in the database. By definition, the Petrosian magnitude for an SDSS galaxy is determined from the total flux inside twice the $r^*$-band Petrosian radius. We determined the total flux within this aperture and computed the zeropoint needed to recover the Petrosian magnitude in the SDSS database. We therefore still rely on the existing SDSS photometric calibration (Fukugita et al. 1996; Hogg et al. 2001; Smith et al. 2002) for the accuracy of our photometry.}.

We chose to measure our morphological parameters at fixed isophotal thresholds of $g^*=24.5$ mag arcsec$^{-2}$ and $i^*=24.0$ mag arcsec$^{-2}$ for the $g^*$ and $i^*$ images, respectively. These isophotal levels were intentionally chosen to be rather bright, for two reasons: Firstly, we wanted the isophote in question to be reachable for all galaxies in the dataset, even those imaged under comparatively poor conditions. Secondly, we wanted to ensure that the isophote in question remained attainable in the rest frame under moderately deep imaging of high-$z$ galaxies with HST, so that the measurements in this paper might be compared with those for high redshift galaxies in a future paper.

\subsection{The relationship between Gini coefficient and central concentration.}

\begin{figure*}[htb]
    \centering
    \includegraphics[width=4.5in,angle=0]{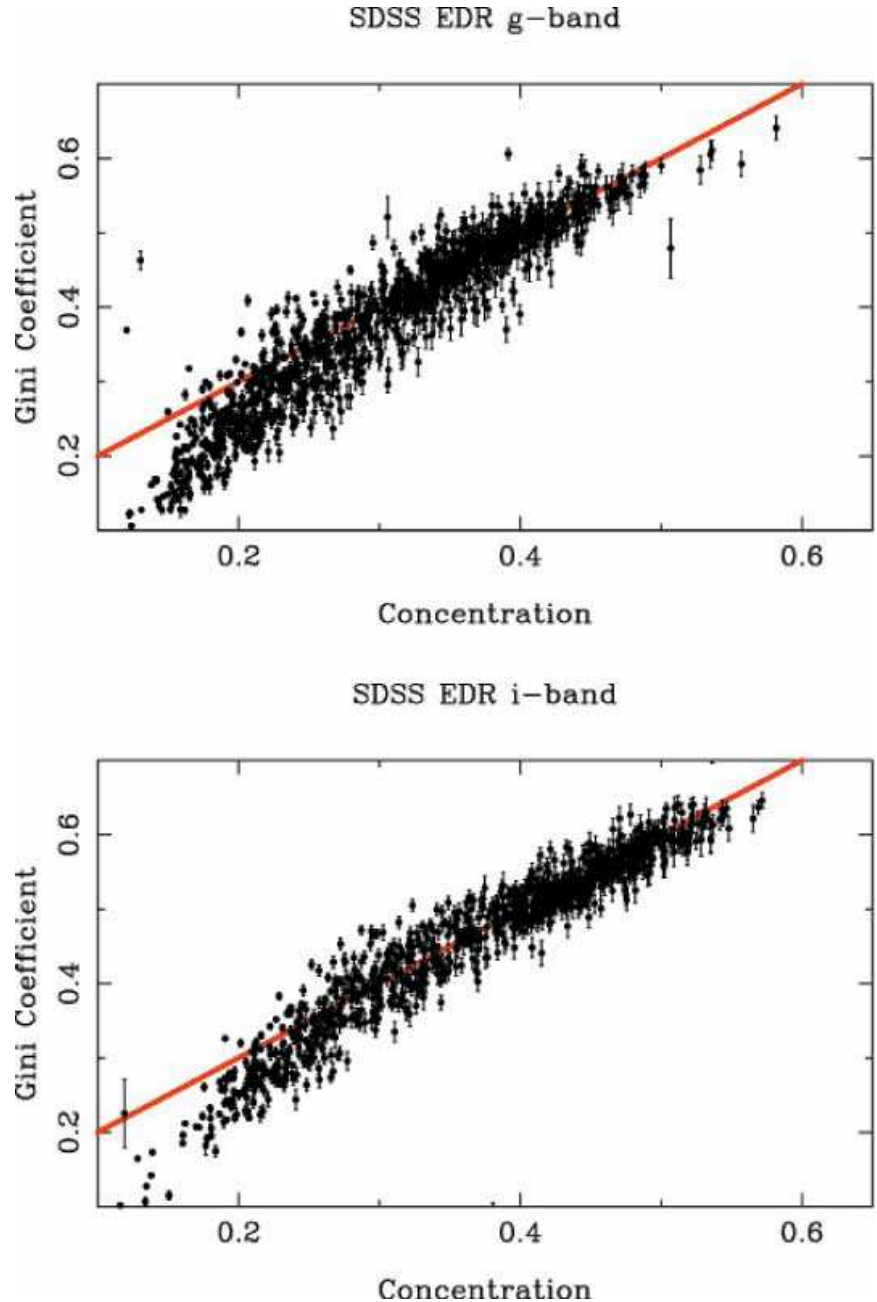}
    \caption{Gini coefficient vs. central concentration for the $g$-band sample (top) and for the $i$-band sample (bottom). The solid line corresponds to unity slope. While the overall slope of the distributions remains similar in both bands, the \gb~ sample exhibits slightly greater scatter and more high-$G$, high-$C$ systems are seen in the \ib. Note that the galaxies span a broad range of morphologies, from pure disk systems at low-$C$ to highly centrally concentrated pure $R^{1/4}$-law elliptical galaxies at high-$C$.}
    \label{fig:GBandAndIBandGiniVsC}
\end{figure*}  

\begin{figure*}[htb]
    \centering
    \includegraphics[height=6.2in,angle=0]{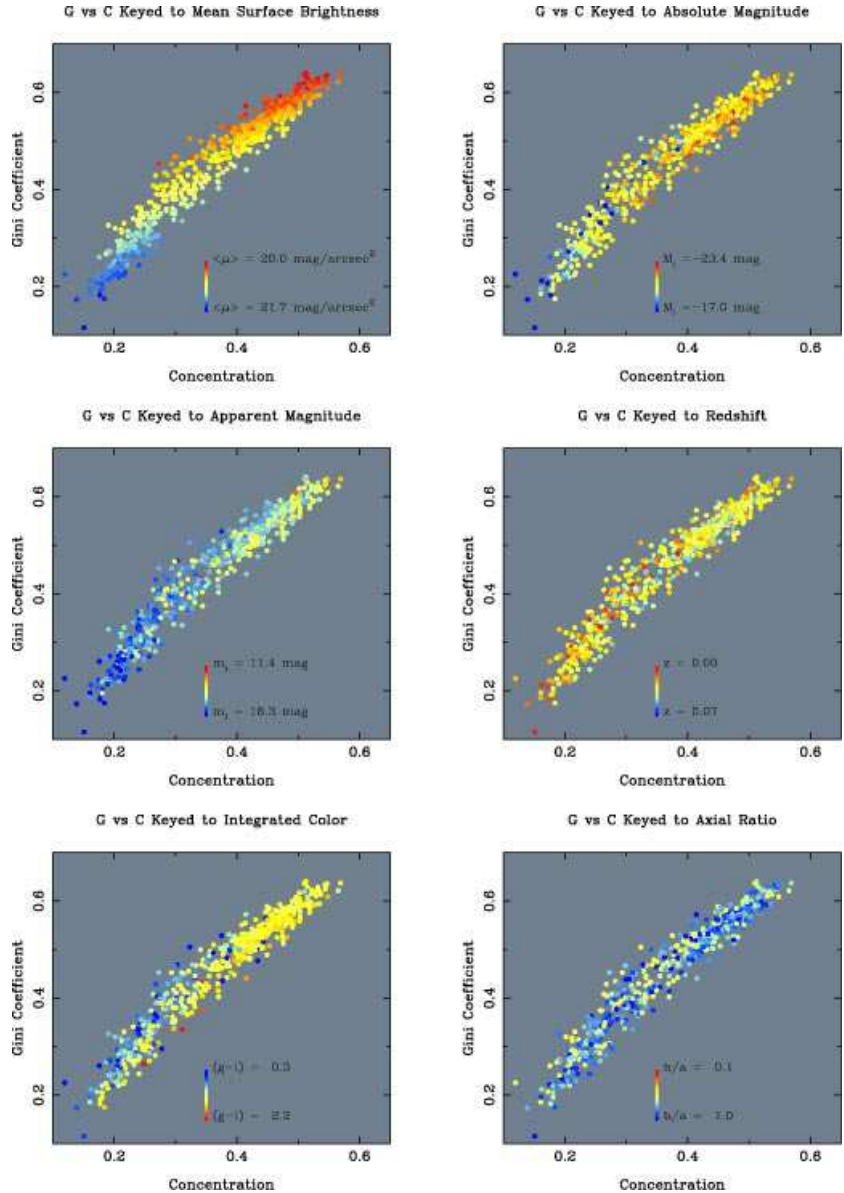}
    \caption{\small An investigation into the possible sources of scatter in the relationship between Gini coefficient and central concentration for the $i$-band sample. Symbol colors in each panel are keyed to a different measurement, with the range of each measurement indicated by the color bar shown as an inset in each panel. Moving clockwise from the top left, panels are keyed to: (a) mean surface brightness, (b) absolute magnitude, (c) redshift, (d) axial ratio, (e) integrated color, (f) apparent magnitude. The strongest systematic source of scatter in $G$ vs $C$ is clearly mean surface brightness, though weak trends are also seen as a function of other quantities. See text for details. Note that error bars have been omitted for the sake of clarity. }
    \label{fig:Everything}
\end{figure*}  

\begin{figure*}[htb]
    \centering
    \includegraphics[width=7in]{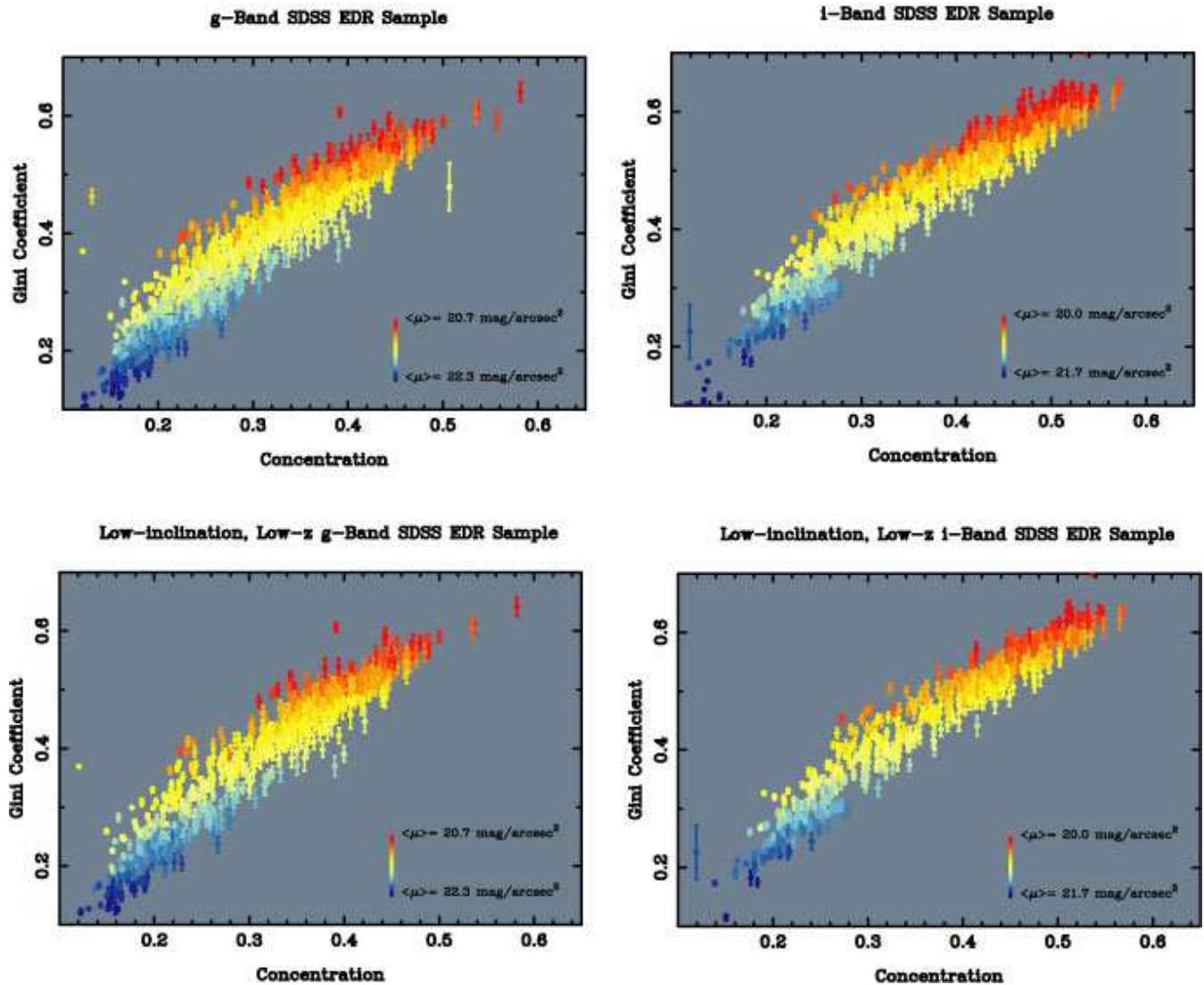}
    \caption{Further investigation into the possible sources of scatter in the relationship between Gini coefficient and central concentration. The left column shows results for the \gb~sample, with the right sample showing results \ib. Each panel shows $G$ vs $C$ with symbol colors keyed to mean surface brightness. The top row corresponds to the full galaxy sample, while the bottom row shows a ``clean'' subset of galaxies with axial ratios ${b\over a}>0.5$ and $z<0.05$ (bottom). Note the systematic trend in the scatter as a function of surface brightness seen in each panel. Little difference is seen between the full sample and the subset.}
    \label{fig:GiniVsCKeyedToSB}
\end{figure*}  

\begin{figure*}[htb]
    \centering
    \includegraphics[width=7in]{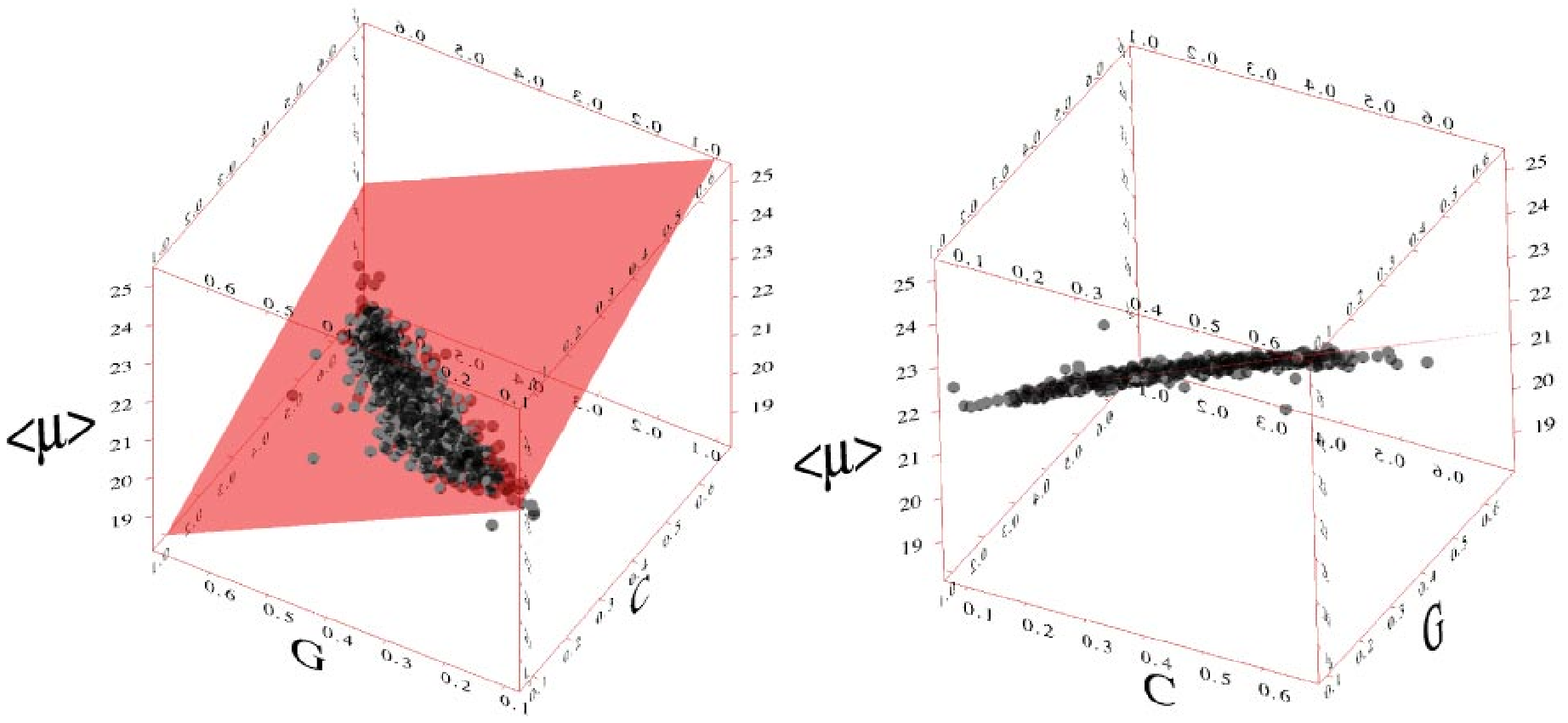}
    \caption{\small Distribution of the \gb~sample in \cgm~ parameter space.The best-fit plane determined from a principial component analysis is shown as a transparent red surface. The projection at left is shown with the coordinates rotated to euler angles \{60$^\circ$, 0$^\circ$, -45$^\circ$\} and illustrates the scatter in the points, while the projection at right shows the best-fit plane edge-on with euler angles  \{-27.5$^\circ$, -1.3$^\circ$, -53.7$^\circ$\}. Note the remarkable thinness of the projected surface in the edge-on view. See text for details}
\label{fig:gband3d}
\end{figure*}

Figure~\ref{fig:GBandAndIBandGiniVsC} illustrates the relationship between the Gini coefficient  and central concentration for both the \gb~ and \ib~images in our sample. Error bars on $G$ ($\pm1\sigma$) are shown, and were determined from 100 bootstrap replications (Efron \& Tibshirani 1993) of the pixel distribution for each galaxy. Bootstrap replication cannot be used to estimate the errors on $C$, and no error bars on $C$ are shown. However, on the basis of the Monte Carlo simulations described in Abraham et al. (1994) we expect error bars on $C$ to be approximately similar in size to the error bars on $G$ (and this expectation will be reinforced by results presented below in \S\ref{sec:plane}). The effects of seeing on these measurements is also small, because even at the limits of our data the galaxies are large relative to the seeing disk. By convolving synthetic point-spread functions against a set of 10 objects representative of the faintest galaxies in our sample, we estimate that in the extreme case of moving from an object with the best seeing in the sample to the worst seeing in the sample, faint galaxies studied here would have $G$ altered systematically downward by about the same size as the $1\sigma$ random error bar shown in the figure. 

The straight line in Figure~\ref{fig:GBandAndIBandGiniVsC} corresponds to a slope of unity and a $y$-axis intercept of 0.1 (i.e. a given galaxy has a slightly greater value of $G$ than $C$). There is clearly a very strong correlation between $G$ and $C$ in both \gb~ and \ib.  The slope of the $G$ vs. $C$ relation remains similar in both bands. The trend is linear with weak curvature at the low-$C$, low-$G$ end, and a slope near unity at the high-$C$, high-$G$ end. More high-$G$, high-$C$ systems are seen in the \ib, as expected from the increased prominence of bulges relative to disks as one moves further into the red. We conclude on the basis of this figure that, on the whole, {\em $G$ makes an excellent substitute for $C$ and relies on fewer basic assumptions about the shape of the galaxy being measured.}

Another obvious feature of these plots is that the \gb~ images exhibit somewhat greater scatter relative to the \ib~ images about the line of unit slope. The crucial point to note, however, is that in both bands the scatter is far larger than the errors on individual measurements of $G$. Therefore at least one additional parameter is needed to explain the relationship between the Gini coefficient and central concentration. To learn more about the nature of the scatter in the $G$ vs. $C$ relationship, Figure~\ref{fig:Everything} presents a multi-panel plot of $G$ vs. $C$ parameter space in the \ib~ with plot symbol colors keyed to a range of intrinsic physical quantities (mean surface brightness, absolute magnitude, integrated color) and possible sources of bias (redshift, apparent axial ratio, apparent magnitude). This figure demonstrates three important points: 

\begin{enumerate}

\item The scatter in the $G$ vs. $C$ relationship is not dominated by systematic biases in the sample selection, redshift distribution,  or galaxy inclination. 

\item The weak trends we expect to see as a function of apparent magnitude and color, which are described below, are in fact seen. Objects with very low central concentration are expected to be very late-type and therefore mostly blue and faint.  One therefore expects a gradient in color and apparent magnitude, with mostly blue galaxies at low $C$ and mostly red galaxies at high $C$. Somewhat more interesting is the observation that, over the range of $C$ dominated by spirals, the bluest galaxies (at a given concentration) tend to have the highest $G$. This can also be interpreted in a manner consistent with our understanding of how star formation occurs in galaxies: At a given concentration (which is closely connected with bulge-to-disk ratio), the more unequal the distribution of starlight (i.e. the more irregular the structure), the more star-formation is occurring.

\item There is a rather obvious and very surprising trend with mean surface brightness that would seem to explain nearly all the scatter in the $G$ vs. $C$ plane.  If  this scatter is really a function of only a single parameter, then all galaxies in nearby regions of the Universe (regardless of morphology, luminosity, mass or star-formation history) might turn out to lie on a single surface in $G$ vs. $C$ vs. $\left<\mu\right>$ parameter space. 

\end{enumerate}

\subsection{A universal structural surface for nearby galaxies}
\label{sec:plane}

\begin{figure*}[htb]
    \centering
    \includegraphics[width=7in]{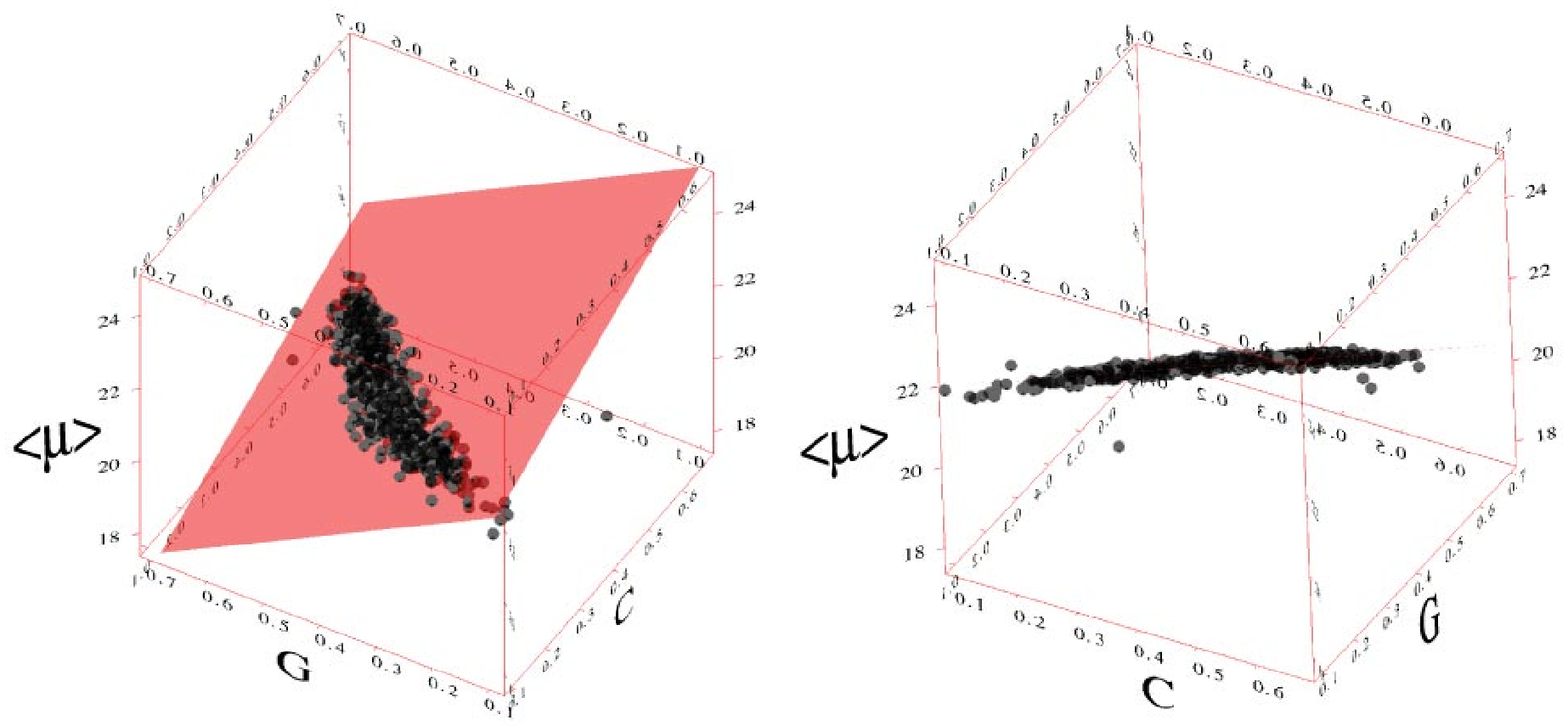}
    \caption{As for the previous figure, except for the \ib~sample. The view at left corresponds to euler anges \{60$^\circ$, 0$^\circ$, -45$^\circ$\}, while the edge-on view at the right correponds to euler angles \{-28.9$^\circ$, -1.8$^\circ$, -52.5$^\circ$\}.}
    \label{fig:iband3d}
\end{figure*}  

Figure~\ref{fig:GiniVsCKeyedToSB} focuses on the parameters explored in the first panel in the previous figure. The figure shows in greater detail the distribution of galaxies in $G$ vs. $C$ vs. $\left<\mu\right>$~ parameter space in both \gb~and \ib, and explores in a little greater detail the sensitivity of the trends seen to redshift and inclination. Symbol colors are once again keyed to mean surface brightness, and cover the full range of this \msb~ spanned by the data. The top row in the Figure shows the full dataset, while the bottom row shows$G$ vs. $C$ vs. $\left<\mu\right>$ for a ``clean'' subset of galaxies with axial ratios $b/a>0.5$ and $z<0.05$. 

The systematic variation with \msb~in the scatter of the $G$ vs. $C$ plane is obvious in data from both filter bands. A given mean surface brightness maps nearly perfectly onto a well-defined stratum on this diagram, so that at any central concentration, the spread in $G$ is a purely monotonic function of mean surface brightness. Each stratum of galaxies delineated by \msb~ appears to follow a linear relationship on the $G$-$C$ plane. Little difference is seen between the full sample and the ``clean'' subset in either band. These trends are seen to be quite robust. The idea that the stratified distribution of galaxies seen in Figure~\ref{fig:GiniVsCKeyedToSB} is the manifestation of the points lying on a single surface in a three dimensional parameter space is investigated in Figures~\ref{fig:gband3d} and~\ref{fig:iband3d}.  These show projections of the galaxy distribution in $G$ vs. $C$ vs. $\left<\mu\right>$ space.  The right-hand panel in each figure shows the parameter space seen from a view obtained by rotating the coordinate system in a manner to illustrate the thinness of the surface defined by the data points. The very marked decrease in the scatter of the points from this perspective confirms that the galaxies lie on a thin surface which is well-fit by a plane. To quantify this, we determined the equations of the minimum variance planes using a principal component analysis. Results from this analysis are shown at superposed on the data points in each of the panels of the figure.  The best fitting planes are described by the following formulae:

\medskip

\gb:
\begin{equation} 
\left<\mu\right> = 22.451 + 5.366\times C - 7.010\times G
\end{equation}  

\ib:
\begin{equation} 
\left<\mu\right> = 22.149 + 5.373\times C - 7.627\times G
\end{equation}  
\medskip

These formulae are useful in two obvious ways. Firstly, they allow us to describe the entire SDSS \gb~ and \ib~samples using two pairs of three numbers: \{22.451, 5.366, 7.010\} (\gb) and \{22.149, 5.373, 7.627\} (\ib). Secondly, they represent convenient formulae that allow one to compute a purely photometric quantity,  $\left<\mu\right>$, from two non-parametric global morphological quantities.

\medskip

\section{DISCUSSION}
\label{sec:discussion}

\subsection{Interpretation of the the characteristic surface}

The finding that nearby galaxies map out a unique surface in a parameter space constructed from $G$, $C$ and $\left<\mu\right>$ is surprising. Our experience from many studies of local galaxies has generally been that the diversity of galaxy ages, masses, morphologies and star formation histories results in scatter diagrams when the physical parameters for these galaxies are plotted against each other, unless carefully defined subsamples are first taken. The mapping of all nearby galaxies onto a single surface in this particular parameter space regardless of variations in morphology and other obvious differences between the galaxies seems an unexpected link between all these systems.  How much of this effect is simply due to internal correlations between $G$, $C$, and $\left<\mu\right>$ introduced by our techniques for measuring them? 

\begin{figure*}[htb]
    \centering
    \includegraphics[width=4.5in]{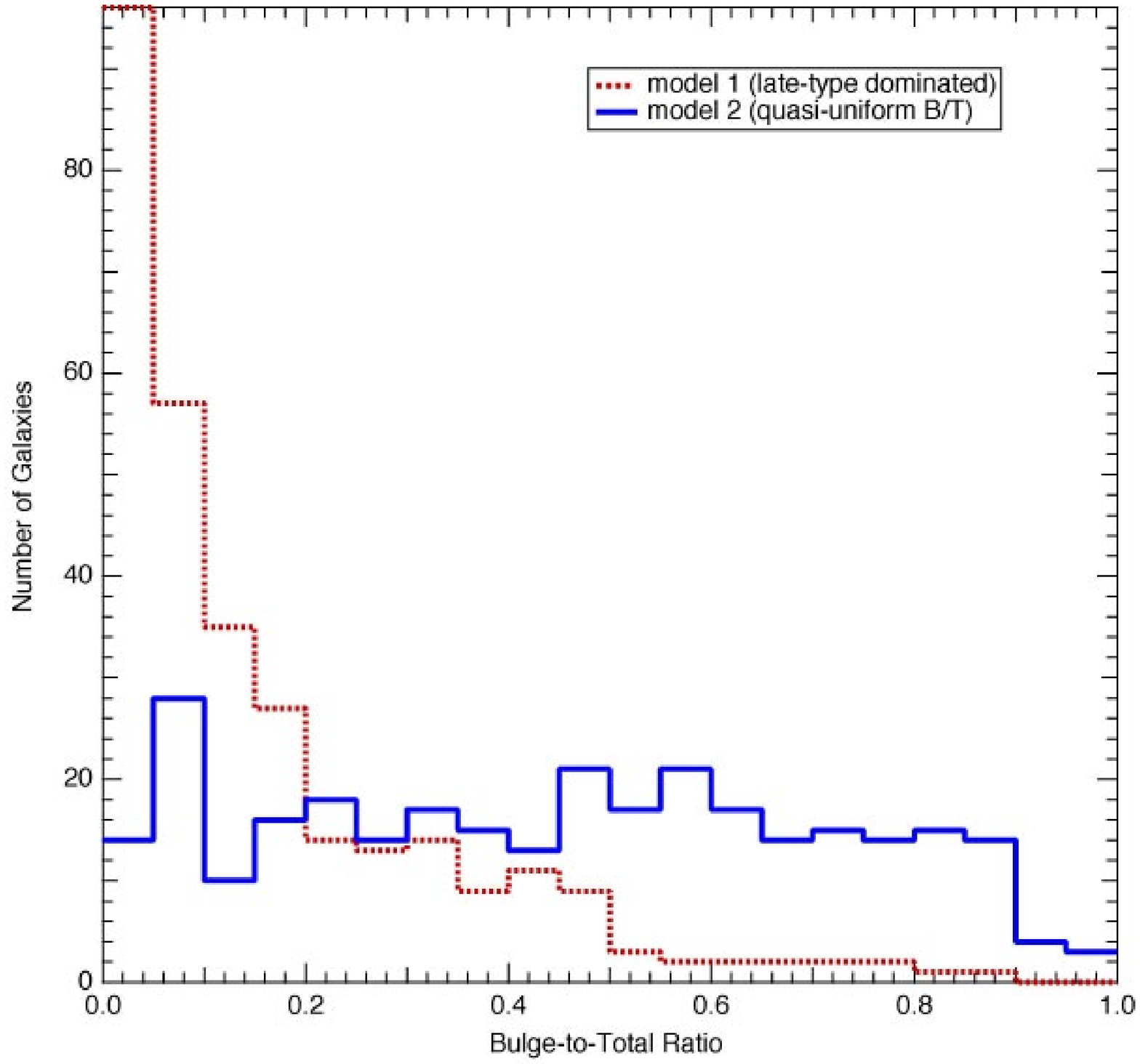}
    \caption{\small Bulge-to-total ratios for the simple models described in the text. These correspond to a distribution dominated by late-type systems (model 1, red dashed line) and a distribution of bulge-to-total ratios that is close to uniform (model 2, blue solid line). As noted in the text, the purpose of these models is to test for degeneracies in \cgm~space, and neither model is intended to closely match the bulge-to-total ratio in the SDSS survey.}
\label{fig:bulgetototal}
\end{figure*}  

\begin{figure*}[htb]
    \centering
    \includegraphics[width=7in]{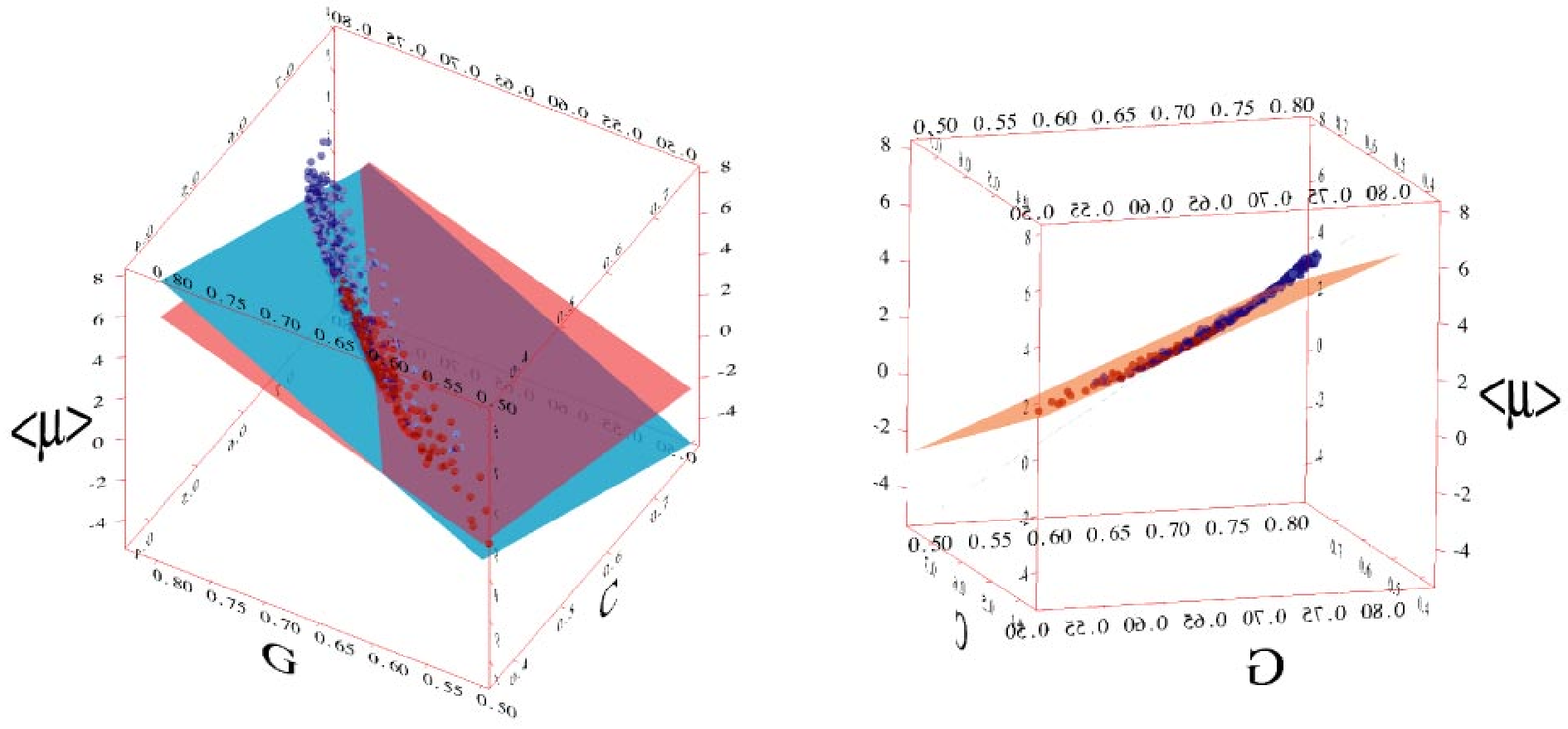}
    \caption{
[Left] Distribution of 600 data points in \cgm~space corresponding to model 1 (disk-dominated) and model 2 (quasi-uniform bulge-to-total ratio). Model 1 is shown in red and model 2 is shown in blue. Note that the zero point of the surface-brightness scaling is arbitrary (to emphasize that measures of $C$ and $G$ are {\em morphological} in nature and do not depend on photometric zero-points). Best-fit PCA planes are also shown as transparent surfaces, keyed to the corresponding colors of the data points. The view is shown  at the standard orientation with euler angles  \{60$^\circ$, 0$^\circ$, -45$^\circ$\} for comparison with the left-hand panels in Figures~\ref{fig:gband3d} and \ref{fig:iband3d}. Note that at the direction of slope of both best-fit planes is different to that seen in the SDSS data (tilting toward the observer in this orientation, while the planes tilt away from the observer in the same orientation in the SDSS observations).  [Right] The same points and surfaces show at left, except viewed edge-on with respect to the plane that is the best fit to model 2. Note the that the surface remains thin but is quite strongly warped, so that the best-fit to model 1 is a poor fit to model 2. The distribution of data points is clearly not degenerate with respect to the underlying models. 
}
    \label{fig:models}
\end{figure*}  

To test whether the characteristic surface is entirely artificial and due to subtle internal correlations in the definitions of our parameters, we generated two sets of simulated galaxies using different distributions of bulge-to-disk ratios (described below)  and determined $G$, $C$ and $\left<\mu\right>$ from these using exactly the same procedure used in analysing the SDSS observations. We emphasize that the toy models we are about to describe are not intended to in any way model the real SDSS data --- our purpose is simply to ask whether very different models converge to the observed surfaces in $G$, $C$ and $\left<\mu\right>$-space because of some underlying fundamental degeneracy. For simplicity, each family of galaxies was described as an exponential disk with an exponential bulge, with one set of models having a distribution of bulge-to-total flux ratios that is skewed heavily to late type galaxies while the other set has a distribution of bulge-to-total flux ratios that is nearly uniform. The distributions of bulge-to-total flux ratios for our models are shown in Figure~\ref{fig:bulgetototal}. The corresponding distributions in $G$ vs. $C$ vs. $\left<\mu\right>$  space are shown in Figure~\ref{fig:models}, along with best-fit planes to the simulated data determined using principal components analysis. 

Several interesting things can be gleaned from Figure~\ref{fig:models}. Firstly, the thinness of the surface \cgm-space should be viewed as somewhat artificial, since quite thin distributions in multi-parameter parameter space are also recovered from both sets of simulations with totally different bulge-to-total flux distributions. However, the shape of the surface is clearly a strong function of the assumed bulge-to-total flux ratio, differing not only for the two simulations when compared against each other, but also differing markedly from the observed data. For example, the direction of slope (tilting toward the observer in this orientation) of both best-fit model planes in the left-hand panel of  Figure~\ref{fig:models} is different to that seen in the SDSS data (tilting away from the observer when viewed at the same orientation). Each model seems to result in points that can be reasonable well fitted to a plane, but when both models are combined the fit to the plane becomes quite poor, as seen in the right hand panel of the figure. We conclude then that the topology and orientation (but not necessarily the thinness) of the characteristic surface is not degenerate. Therefore the \cgm~ parameter space may provide a useful tool for quantifying the overall structural distribution of galaxies, and we very tentatively suggest that the flatness of the surface seen in our observed data might be associated with an underlying homology in the underlying galaxy distributions. 

\subsection{Sensitivity of results to different methods used to compute central concentration}

\begin{figure*}[htb]
    \centering
    \includegraphics[width=5.5in,angle=0]{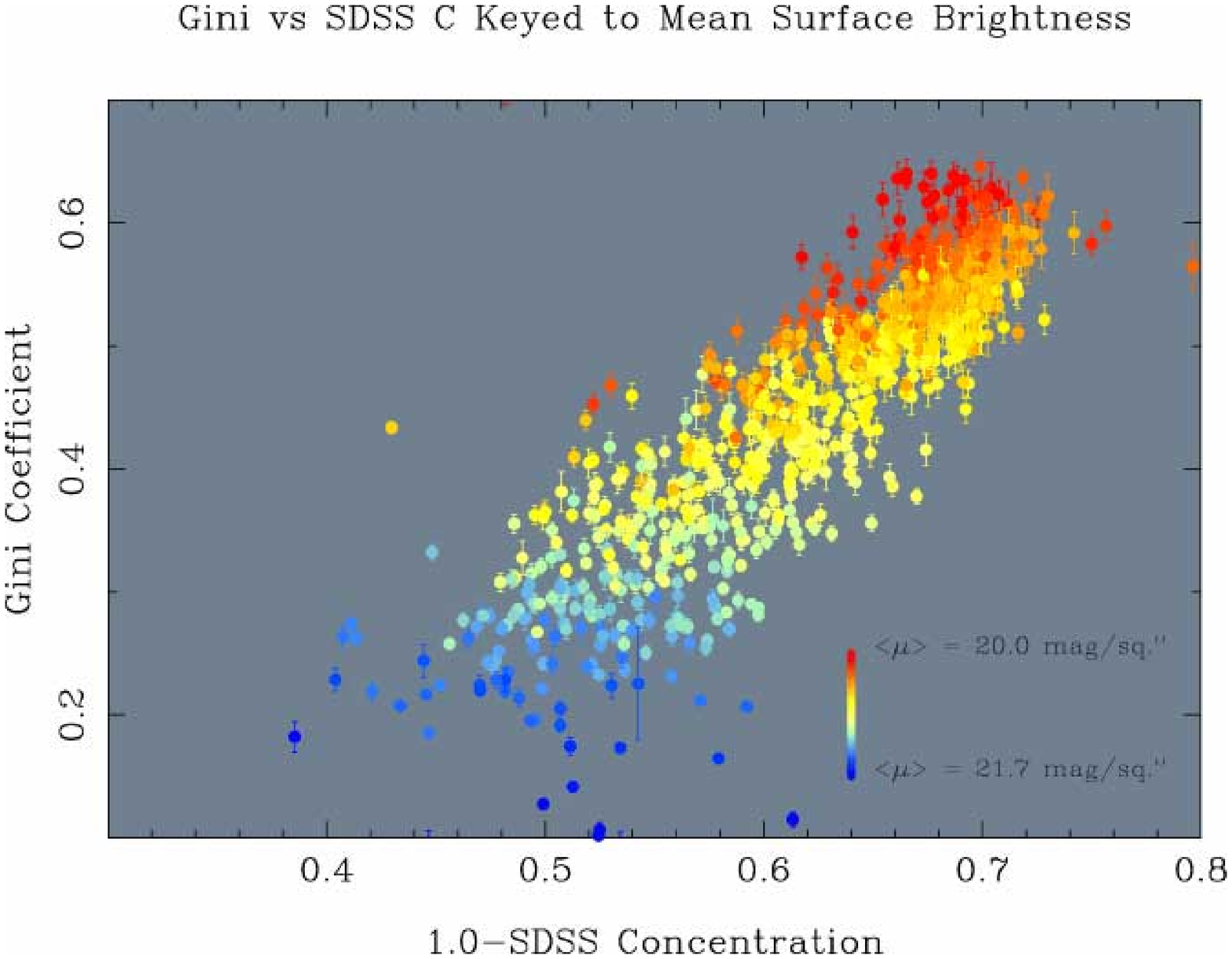}
    \caption{Gini coefficient plotted as a function of central concentration recorded in the SDSS database. Symbols are keyed to mean surface brightness. The relative effectiveness of different measures of central concentration can be determined by comparing this figure with figure \ref{fig:GiniVsCKeyedToSB}.}
    \label{fig:CVsSDSSC}
\end{figure*}  

There is at present no standard definition of central concentration that is universally used in the analysis of galaxy images. The plots presented here have been based on $C$ as determined from the moment-based technique described in Abraham et al. (1994). This concentration measure is computed from flux ratios within elliptical apertures determined from second-order image moments. It is interesting to consider whether any of the results presented in this paper depend on the precise definition of central concentration we have used. More specifically, could the results obtained here also be found using the alternative definition of central concentration determined by the SDSS Photo pipeline? The SDSS concentration indices are based upon ratios of Petrosian radii determined using circular apertures, and have the benefit of distance independence but the disadvantage of not accounting for the effects of galaxy inclination.  The top panel of Figure \ref{fig:CVsSDSSC} shows $C$ vs. $G$ vs. $\left<\mu\right>$~parameter space determined using SDSS Petrosian concentration indices. The trends presented in earlier figures remain, but they are washed out, and the stratification of scatter as a function of surface brightness is less clear. We conclude that the SDSS concentration indices are certainly useful, but retain a decided preference for our own definition of $C$ for use in studying very bright galaxies. The galaxies studied in the present paper span only a tiny redshift range, so the major benefits of Petrosian-based SDSS concentration indices do not yet outweigh the noise added by the use of circular apertures.  In any case, the important point is that our general results do not appear to depend strongly on the precise definition of central concentration used in this paper.

\subsection{Applications of the characteristic surface to studies of galaxy evolution}

A promising avenue for using the ideas presented in this paper to quantify galaxy evolution is to use the narrow surface in \cgm~parameter space as a fiducial benchmark. Placing galaxies in the relevant parameter space can be done without dynamical measurements (which are difficult to obtain at high redshifts). Position relative to the surface defined at low redshift might then be used to calibrate evolution in the same manner that position relative to the local Tully-Fisher and Fundamental Plane relations have provided robust calibrators for evolution in size, surface brightness, and mass-to-light ratio (Vogt et al 1997; Vogt 2001; van Dokkum et al. 2001; van Dokkum \& Franx 2001). Evolutionary changes could manifest themselves as either a displacement of points on the surface (if we assume that all high-redshift galaxy forms are represented somewhere within the SDSS EDR, with a changing mix but no fundamentally new forms represented at high redshifts), or as a distortion of the surface itself. Observation of the latter effect would provide concrete evidence for new classes of high-redshift systems that have since disappeared. In any case, reproducing this surface poses a concrete and simple test of galaxy formation models, with obvious application to calibrating $n$-body/hydro simulations.

As redshift increases, measurements of $C$ become ever less meaningful as the galaxy population becomes increasingly distorted. However, since measurements of $G$ and $\left<\mu\right>$ are possible for galaxies of arbitrary shape, we expect that mapping the loci of galaxy distributions in the parameter space defined by these two quantities alone will prove interesting. A particularly useful application would be to map out strongly lensed background galaxies onto such a parameter space. Such a program might allow morphological work to be extended to the extremes of redshift space, though of course the important issue of synchronization to a uniform rest wavelength and accounting for cosmological dimming over a range of redshifts would still need to be addressed. Extension of the present study to higher redshifts will be the subject of future work.

\section{CONCLUSIONS}
\label{sec:conc}

In this paper we have introduced the Gini coefficient as a new tool for characterizing the morphologies of galaxy images. This statistic is a robust measure of dispersion that is confined to the unit interval. To a first approximation the Gini coefficient of a galaxy image can be viewed as a type of generalized concentration index that does not rely on any underlying symmetry in a galaxy, and which does not even require a galaxy's center to be well-defined.  While the Gini coefficient does not measure exactly the same thing as a conventional concentration index,  it does appear to be sufficiently close to concentration that it can probably be used as a meaningful substitute for concentration in studies of high-redshift systems where many galaxies are expected to be distorted or peculiar. For local galaxies the Gini coefficient should be viewed as a useful supplement to, and not a replacement for, measures of concentration.

On the basis of an analysis of 930 $g^\star<16$ mag nearby galaxies from the {\em Early Data Release} of the {\em Sloan Digital Sky Survey}, we have quantified the relationship between the Gini coefficient and central concentration of galaxies. We find that the scatter in the tight $G$ vs. $C$ relationship is surprisingly well-correlated with mean surface brightness $\left<\mu\right>$, and show that nearby galaxies occupy a thin two-dimensional surface embedded within the three-dimensional $G$ vs. $C$ vs. $\mu$ space, and present equations describing the best-fit plane describing this surface.   By associating each galaxy sample with the equation of this plane, we encode the morphological composition of the entire SDSS \gb~sample as \{22.451, 5.366, 7.010\}, and the \ib~sample as: \{22.149, 5.373, and 7.627\}. We have demonstrated that a thin surface in this parameter space is a generic feature of simulated galaxy distributions, although the detailed topology and orientation of the surface depends on the structural distribution of the galaxies in the sample. By defining a simple point of reference for evolutionary studies of changing morphology, position relative to the universal surface in $G$ vs. $C$ vs. $\mu$ space provides an objective zero-point defining the structural distribution of nearby galaxies. In this context, it will be interesting to learn how this~ surface is shifted (or distorted) as a function of redshift and environment.

\acknowledgments
\centerline{\em Acknowledgments}

\noindent It is a pleasure to thank Karl Glazebrook, Henk Hoekstra, Howard Yee and Richard Ellis for useful suggestions. We are grateful to the referee, Matt Bershady, for many useful comments which greatly improved this paper. Timely assistance in the production of 3D figures was received from the technical staff at Wavemetrics Inc., and is gratefully acknowledged. Finally, we would also like to thank the many members of the SDSS team for their years of hard work and dedication in putting the survey together.

Funding for the creation and distribution of the SDSS Archive has been provided by the Alfred P. Sloan Foundation, the Participating Institutions, the National Aeronautics and Space Administration, the National Science Foundation, the U.S. Department of Energy, the Japanese Monbukagakusho, and the Max Planck Society. The SDSS Web site is http://www.sdss.org/.

The SDSS is managed by the Astrophysical Research Consortium (ARC) for the Participating Institutions. The Participating Institutions are The University of Chicago, Fermilab, the Institute for Advanced Study, the Japan Participation Group, The Johns Hopkins University, Los Alamos National Laboratory, the Max-Planck-Institute for Astronomy (MPIA), the Max-Planck-Institute for Astrophysics (MPA), New Mexico State University, Princeton University, the United States Naval Observatory, and the University of Washington.


\end{document}